\documentclass[aps, prl, 10pt, superscriptaddress, twocolumn]{revtex4-1}
	\usepackage[bookmarks=false,linkcolor=blue,urlcolor=blue,colorlinks,citecolor=blue]{hyperref}

	\usepackage{graphicx}
	\usepackage{color}
	\usepackage{amsmath}
	\usepackage[normalem]{ulem}
	\usepackage[toc,page]{appendix}
	\usepackage{amsfonts}
	\usepackage{amssymb}

\newcommand{\be}{\begin{equation}}
\newcommand{\ee}{\end{equation}}
\newcommand{\p}{\partial}

\newcommand{\la}{\label}
\newcommand{\bea}{\begin{eqnarray}}
\newcommand{\eea}{\end{eqnarray}}

\begin{document}

\title{\begin{flushright}\vspace{-1in}
			\mbox{\normalsize }
		\end{flushright}
	Chiral Topological Elasticity and Fracton Order 
	 \vskip 20pt
	 }

\author{Andrey Gromov}
\affiliation{Brown Theoretical Physics Center and Department of Physics, Brown University, Providence, Rhode Island 02912, USA}

\date{\today}

%%%%%%%%%%%%%%%%%%%%%%%%%%%%%%%%%%%%%%%%%%%%%%%%%%%%%%%%%%%%%%%%%%%%%%%%%%%%%%%%%%%%%%%%%%%%%%%
\begin{abstract}
We analyze the ``higher rank'' gauge theories, that capture some of the phenomenology of the Fracton order. It is shown that these theories lose gauge invariance when arbitrarily weak and smooth curvature is introduced. We propose a resolution to this problem by introducing a theory invariant under area-preserving diffeomorphisms, which reduce to the ``higher rank'' gauge transformations upon linearization around a flat background. The proposed theory is \emph{geometric} in nature and is interpreted as a theory of \emph{chiral topological elasticity}. This theory exhibits some of the Fracton phenomenology. We explore the conservation laws, topological excitations, linear response, various kinematical constraints, and canonical structure of the theory. Finally, we emphasize that the very structure of Riemann-Cartan geometry, which we use to formulate the theory, encodes some of the Fracton phenomenology, suggesting that the Fracton order itself is \emph{geometric} in nature. 
\end{abstract}

%%%%%%%%%%%%%%%%%%%%%%%%%%%%%%%%%%%%%%%%%%%%%%%%%%%%%%%%%%%%%%%%%%%%%%%%%%%%%%%%%%%%%%%%%%%%%%%

\maketitle

%%%%%%%%%%%%%%%%%%%%%%%%%%%%%%%%%%%%%%%%%%%%%%%%%%%%%%%%%%%%%%%%%%%%%%%%%%%%%%%%%%%%%%%%%%%%%%%

%%%%%%%%%%%%%%%%%%%%
\paragraph{Introduction.---} 
%%%%%%%%%%%%%%%%%%%%
A new exciting type of order, dubbed Fracton order, was recently introduced \cite{haah2011local, yoshida2013exotic, vijay2015new, vijay2016fracton,williamson2016fractal}, and has attracted a lot of attention \cite{pretko2017subdimensional, ma2017fracton, pretko2017higher,devakul2017correlation,ma2017topological,petrova2017simple, slagle2017quantum}. Particles with restricted mobility, fractal operators and unusual ground state degeneracy are among the exotic features of the Fracton models. These models appear to remain gapped in thermodynamic limit and support a sub-extensive number of groundstates. It was argued that some of the Fracton phenomenology is captured by the (gapless) effective ``higher rank'' gauge theories (introduced in \cite{xu2006novel}) \cite{pretko2017subdimensional}, where the degrees of freedom are ``higher rank gauge fields''. The Gauss law constraints in these models led to the particles with restricted mobility. A particularly simple set of (chiral) models was introduced in two spatial dimensions \cite{pretko2017higher, prem2017emergent}, although the construction of a microscopic model realizing these chiral phases remains an open problem. Symmetric tensor degrees of freedom have also recently appeared in quantum Hall physics \cite{you2014theory, maciejko2013field,PhysRevLett.119.146602, gromov2017Bimetric, cappelli2016multipole}.

In this Note we critically revisit the higher rank gauge theories. We start by showing that higher rank gauge symmetry breaks down in a weakly curved space. We interpret this fact as a fundamental inconsistency of these theories as theories with internal gauge symmetry group. We propose a resolution of this inconsistency by interpreting the higher rank gauge symmetry as a spatial symmetry under area-preserving diffeomorphisms. Armed with this interpretation we construct a covariant theory that reduces to the two dimensional higher rank models upon linearization around the flat background.

We interpret the proposed effective theory as a theory of \emph{chiral topological elasticity}. This theory is topological in that it does not require ambient metric to define the classical Lagrangian on an arbitrary manifold, but its symmetry group is \emph{geometric} in nature, as it is related to spatial translations. The effective theory describes the quantum elastic medium  formed on top of a classical lattice. The excitations in the medium are quantum dislocations with the Burgers vector that can be a fraction of the lattice vector of the classical lattice. The dislocations satisfy a (non-linear) glide constraint and move along one-dimensional sub-manifolds. Glide constraint follows from the (non-linear) ``volume'' conservation law. The disclinations are high energy immobile excitations that can only move by creating dislocations along the way \cite{LandauLifshitz-7}. Relation between Fracton order and elasticity was very recently discussed in \cite{PretkoPRL2018}.

%%%%%%%%%%%%%%%%%%%%%%
\paragraph{``Higher rank'' gauge theory.---} 
%%%%%%%%%%%%%%%%%%%%%%

We start with a brief review of the effective theory for the $2$D chiral Fracton order proposed in \cite{pretko2017higher, prem2017emergent}. The degrees of freedom are described by a rank-$2$, symmetric, traceless gauge field $a_{ij}$, and a scalar Lagrange multiplier $\chi$. The action is given by \cite{pretko2017higher, prem2017emergent}
\be\la{eq:Paction}
 S_{\rm 2} = \frac{k}{4\pi}\int dt d^2x\Big( 2\chi  \epsilon^{ij} \p_i \p_k a_{j}{}^k - \epsilon^{ij} a_{ik} \p_0 a_j{}^k\Big)\,,
\ee
where $i,j,\ldots=1,2$.
The action \eqref{eq:Paction} is invariant under the gauge transformations
\be \la{eq:gaugeT}
\delta a_{ij} =  \left[\p_i \p_j - \frac{1}{2}\delta_{ij}\Delta\right]\alpha\,, \qquad \delta \chi = \dot \alpha\,,
\ee
where $\alpha$ is the scalar gauge parameter.

We would like to show that the higher rank gauge theory becomes inconsistent in a weakly curved space. We will assume that $a_{ij}$ is a true rank-$2$ tensor \footnote{This must be the case since the r.h.s. of \eqref{eq:gaugeT} is a rank-$2$ tensor.}. Rank-$2$ tensor is a two-index object that transforms under a coordinate change $x^i \rightarrow x^i + \xi^i(x)$ as follows
\be\la{eq:diff2}
\delta a_{ij} = \xi^k \p_k a_{ij} + a_{ik}\partial_j \xi^k + a_{jk}\partial_i \xi^k\,.
\ee
Clearly, \eqref{eq:Paction} is not invariant under \eqref{eq:diff2}. To restore the invariance one has to replace all partial derivatives by the covariant derivatives $\nabla_i$, and contract all indices with the ambient metric $g_{ij}$. For the action \eqref{eq:Paction} we obtain
\be\la{eq:PactionDiff}
 S_{\rm 2} = \frac{k}{4\pi}\int dt d^2x \sqrt{g}\Big( 2\chi  \epsilon^{ij} g^{lk}\nabla_i \nabla_k a_{jl} - \epsilon^{ij}g^{lk} a_{ik} \p_0 a_{jl}\Big)\,.
\ee
This action must be supplemented with the covariant form of the gauge transformation $\delta a_{ij} = \left[\nabla_i \nabla_j - \frac{1}{2}g^{ij}\nabla_i \nabla_j\right]\alpha$. Under a time-independent gauge transformation the second term in \eqref{eq:PactionDiff} is  invariant, while the first term is not
\be\la{eq:varPS}
\delta S_{\rm 2} \sim \int dt d^2x \sqrt{g} \alpha \epsilon^{ij} \left(\nabla_i R\right) \left(\nabla_j \chi\right) \neq 0\,,
\ee
where $R$ is the Ricci curvature. To derive \eqref{eq:varPS} we have used the definition of the Riemann tensor $\left[\nabla_i,\nabla_j\right] v^k = R^k{}_{lij} v^l$ and the following explicit $2$D formula \cite{nakahara2003geometry}  $R_{ijkl} = \frac{R}{2} \left(g_{ik} g_{jl} - g_{il}g_{jk}\right)$. Since the integral in \eqref{eq:varPS} does not generally vanish, we conclude that the ``higher rank gauge symmetry'' is broken in curved space \footnote{The reader may be tempted to argue that our observation implies that higher rank gauge theory is only well-defined in flat space, because it is related to the microscopic model that is very sensitive to the lattice effects. We believe that the effective gauge theory cannot make sense if adding arbitrarily small and smooth gaussian curvature breaks the gauge invariance. Alternative would imply that the number of degrees of freedom depends on whether the space is curved.}. The Eq.\eqref{eq:varPS} is the first central result of this Note \footnote{Sensitivity of the lattice Fracton models to geometry was remarked in \cite{slagle2017quantum}, where it was found that non-trivial geometry leads to a robust groundstate degeneracy. The relation of this fact to our observation \eqref{eq:varPS} is not clear}. The present argument holds true in \emph{any} dimension and for the Lagrangians that include a Maxwell-type term. This follows from the fact that gauge symmetry is ensured by the commutativity of the partial derivatives in flat space, which certainly breaks down in curved space. In what follows we show that it is possible to circumvent this issue by abandoning the ``higher rank'' gauge theory interpretation of \eqref{eq:Paction}-\eqref{eq:gaugeT}. 

%%%%%%%%%%%%%%%%%%%%%%
\paragraph{Area-preserving diffeomorphisms.---} 
%%%%%%%%%%%%%%%%%%%%%%
We would like to construct an effective theory that has some version of \eqref{eq:gaugeT} as a symmetry, \emph{and} is well-defined on an arbitrary manifold. First, we will formulate the theory and then prove that it reduces to \eqref{eq:Paction}-\eqref{eq:gaugeT} in a particular limit. 

The degrees of freedom will be described by the vielbein field $\hat e^A_\mu$ \cite{carroll1997lecture,nakahara2003geometry}, where $A=1,2$, $\mu=0,1,2$. The action is given by
\be\la{eq:FQE}
S_{\rm CTE} = \frac{k}{4\pi}\delta_{AB} \!\int\! d^3x \epsilon^{\mu \nu \rho} \hat e^A_\mu \hat T^B_{\nu\rho} \equiv \frac{k}{4\pi}\delta_{AB} \!\int\!\hat e^A \wedge \hat T^B\,,
\ee
where $\hat T^A_{\nu\rho}$ is the torsion $2$-form \cite{carroll1997lecture,nakahara2003geometry}. The vielbeins $\hat e^A_\mu$ describe emergent degrees of freedom and not the geometry of ambient space\footnote{Note that $k$ has the dimension of length. As we will argue, $S_{\rm FTE}$ describes a quantum lattice, that will supply us with a length scale --- the lattice constant, which we set to $1$.}.

Under the coordinate change vielbeins transform as
\be\la{eq:diff}
\delta \hat e^A_\mu = \xi^\nu \p_\nu \hat e^A_\mu + \hat e^A_\nu \p_\mu \xi^\nu\,. 
\ee
The action \eqref{eq:FQE} is invariant under these transformations.

We now show that Eqs.~\eqref{eq:FQE}-\eqref{eq:diff} reduce to \eqref{eq:Paction}-\eqref{eq:gaugeT}. Consider the vielbeins of the form
\be
\hat e^A_\mu = \delta^A_\mu + (\epsilon^{AB}\partial_B\chi, -\epsilon^{A}{}_B \delta^{Bi} a_{ij})\,,
\ee
where $\chi, a_{ij}$ should be viewed as small fluctuations around $\delta^A_\mu$. We are going to impose a constraint on  \eqref{eq:diff} by requiring that determinant of the vielbein, $\hat e$, is preserved $\delta \hat e = 0$. This type of transformations is called an area-preserving diffeomorphism (APD). It satisfies
\be
\p_i \xi^i = 0 \qquad \Rightarrow \qquad \xi^i = \epsilon^{ij} \p_j \alpha\,.
\ee
Under the APDs we find the transformation laws \eqref{eq:gaugeT}. In terms of the variables $(\chi, a_{ij})$  \eqref{eq:FQE} reduces to \eqref{eq:Paction}.

What did we accomplish? We have constructed a topological theory, that is well-defined on an arbitrary manifold and that reduces to \eqref{eq:Paction}-\eqref{eq:gaugeT}, upon linearization around a particular background. The ``gauge transformations'' were identified with the subgroup of the diffeomorphisms that preserve the volume element $\hat e$. There is no \emph{internal} gauge symmetry in the problem \footnote{In fact, we believe that it is not possible to reconcile the higher rank gauge invariance with the reparametrisation invariance, unless the former is interpreted as a subset of the latter.}. The action \eqref{eq:FQE} is topological in that it is independent of the metric of the ambient space\footnote{Upon certain choices this action can be viewed as a Chern-Simons theory for the gauge group of translations.}.

%%%%%%%%%%%%%%%%%%%%%%
\paragraph{Chiral Topological Elasticity.---} 
%%%%%%%%%%%%%%%%%%%%%%
We will argue that the theory \eqref{eq:FQE} is a quantum theory of elasticity with fractionalized excitations, for $k>1$. We take inspiration from the geometric formulation of elasticity and defects \cite{kleinert1989gauge, katanaev1992theory, katanaev2005geometric, lazar2000dislocation,lazar2011fundamentals,malyshev2000t}. The simplest such formulation involves \emph{teleparallel} \cite{hehl2007elie,kleinert2011new} geometry, \emph{i.e.} curvature-free geometry with torsion. The geometric objects involved are the vielbeins $\hat e^A_\mu$ and the torsion $2$-form $\hat T^A$. The two are related by the Cartan structure equation \cite{carroll1997lecture,nakahara2003geometry}
\be\la{eq:Torsion}
d \hat e^A = \hat T^A\, \quad \Rightarrow \quad S_{\rm CTE} = \frac{k}{4\pi} \delta_{AB} \!\int\! \hat e^A \wedge  d \hat e^B\,.
\ee
The classical phase space is spanned by the torsionless vielbeins satisfying $\epsilon^{ij}\p_i\hat e^A_j = \epsilon^{ij}\hat T^A_{ij} = 0$. The term \eqref{eq:FQE} can be generated by coupling a chiral matter (such as massive Dirac fermion) to the torsional geometry \cite{hughes2012torsion, barkeshli2012dissipationless} and then allowing the geometry to fluctuate. The action \eqref{eq:FQE}, with $k=1$, appears to capture the essential features of the topological mechanics discussed in \cite{paulose2015topological}.

To get further insight into the theory \eqref{eq:Torsion} we turn on the background geometry, described by another set of vielbeins, $e^A_\mu$ (no ``hat''). The action takes form
\be\la{eq:FQEsource}
S_{\rm CTE} = \frac{k}{4\pi}\delta_{AB} \!\int\! \hat e^A \wedge d \hat e^B - \frac{1}{2\pi}\delta_{AB} \!\int\! e^A \wedge d \hat e^B\,.
\ee
This theory is invariant under the diffeomorphisms performed on $\hat e^A$ and $e^A$ simultaneously.

Invariance of the action under the general diffeomorphisms implies a local (flat space) momentum conservation law
\be\la{eq:momentum}
\partial_\mu \sigma_A^\mu = 0\,, \qquad \sigma_A^\mu = \frac{1}{2e}\frac{\delta S}{\delta e^A_\mu}\,,
\ee
where $\sigma^0_A = P_A$ is identified with momentum density and $\sigma^i_A$ is the momentum current and $e= \det e^A_i$. We can switch between the $A,B,\ldots$ and $i,j,\ldots$ indices using the ambient (flat) vielbeins $e^A_i = \delta^A_i$. The existence of the identification between the internal indices $A$ and manifold indices $i$ is the fundamental difference between the present theory and the usual Chern-Simons theory with internal gauge group. From now on we will freely switch between the indices using $P_i = \delta^A_i P_A, \sigma^i{}_j = \delta^A_j \sigma^i{}_A,$ \emph{etc}.

When restricted to \emph{area-preserving} diffeomorphisms the conservation law becomes
\be
\epsilon^{AB} \partial_A \partial_\mu \sigma_B^\mu = 0\,\quad \Leftrightarrow \quad \epsilon^{AB}\p_A\dot P_B + \epsilon^{AB}\p_A \p_j \sigma_B^j = 0\,.
\ee
Upon linearization this can be re-written as a conservation for a ``density'' $\varrho$ (c.f. Ref.\,\cite{pretko2017subdimensional})\footnote{Upon using \eqref{eq:GB} we can identify $\varrho$ with the Ricci curvature $\hat R^{\rm LC}$}
\be
\dot \varrho + \p_i \p_j J^{ij} = 0\,,
\ee
where we have defined
\be
 \varrho = \epsilon^{ij} \p_i P_j\,, \qquad J^{ij} = \frac{1}{2} \Big[\sigma^i{}_k\epsilon^{kj} + \sigma^j{}_k\epsilon^{ki}  \Big]\,.
\ee
 Some of the Fracton phenomenology, as discussed in \cite{pretko2017subdimensional}, follows from this conservation law. For example, the ``dipole moment'' $D_i$
\be
D_i = \!\int\! d^2x\,\, x_i \varrho = \!\int\! d^2x \,\, x_i \epsilon^{kj} \p_j  P_k  =  \!\int \!d^2x \,\,   \epsilon_i{}^{j}P_j\,
\ee
is conserved. The local dipole moment $d_i = \epsilon_i{}^j P_j$ is perpendicular to the momentum: the dipoles always move perpendicular to their dipole moment.

%%%%%%%%%%%%%%%%%%%%%%
\paragraph{Operator content.---} 
%%%%%%%%%%%%%%%%%%%%%%

Differentiating the action \eqref{eq:FQEsource} with respect to $e^A_\mu$ we find the momentum and momentum current operators (recall that $\hat e^A_\mu$ are dynamical)
\be\la{stress}
\sigma^\mu_A = \frac{1}{2\pi} \epsilon^{\mu\nu\rho} \p_\nu \hat e_{A,\rho}\,, \qquad \p_\mu \sigma^\mu_A \equiv  0\,. 
\ee
Note that the current $\sigma^\mu_A$ is conserved \emph{identically}. Note that Eq.\eqref{stress} can be viewed as the starting point for an elastic version of the particle-vortex duality. Indeed, in the absence of external forces any translationally invariant theory, including elasticity, must satisfy the momentum conservation law \eqref{eq:momentum}. Momentum conservation can be solved exactly by \eqref{stress}. The ``gauge'' freedom in choosing the solution of \eqref{eq:momentum} is formally identical to the local translation invariance \footnote{To see this we decompose the vielbein as $\hat e^A_\mu = \p_\mu x^A -a^A_\mu$. Where $x^A$ is the non-singular part of the vielbein. Then the stress tensor $\sigma^\mu_A$ is determined completely by the singular part of the vielbein $a^A_\mu$. The field $a^A_\mu$ is a gauge field for the local translations $x^A\rightarrow x^A+\lambda^A$ and transforms according to $\delta a^A_\mu = \p_\mu \lambda^A$, so that the vielbein is invariant. This gauge freedom is exactly the freedom in solving \eqref{eq:momentum}} that appears in the gauge approach to elasticity (see Refs.\cite{lazar2000dislocation,lazar2011fundamentals,malyshev2000t}).

\begin{figure}
\includegraphics[width=2.5in]{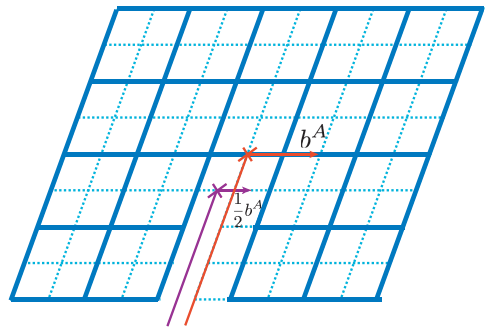}
\caption{Chiral topological elastic medium, $\hat e^A_\mu$, can be visualized on top of a classical lattice that is used to probe the system, $e^A_\mu$. Here the classical background lattice is drawn with solid lines, while the quantum lattice is dashed. The lattice constant of the quantum lattice is twice smaller than the one of the background, corresponding to $k=2$. Dislocations of a quantum lattice carry a fractional (in the units of primitive lattice vectors of the classical lattice) Burgers vector. When $k=1$ two lattices coincide. Another possibility (for $k=1$) is the that the background lattice is actually the \emph{dual} lattice \cite{paulose2015topological}. If $k$ is not an integer then the two lattices are incommensurate.}
\label{mom}
\end{figure}

Consider a background with a singular configuration of torsion, \emph{i.e.} a dislocation at position $x=a(t)$ with Burgers vector $b_A$ \cite{katanaev2005geometric}. Then, by the equations of motion,
\be
\hat T_A = \frac{1}{k} b_A \delta(x-a)\,.
\ee
 The momentum and momentum current localized on a dislocation are
\be\la{eq:currents}
P_A = \frac{ b_A}{2\pi k} \delta(x-a)\,,\qquad \sigma^i_A = \frac{b_A}{2\pi k}  \dot{a}^i \delta(x-a)\,.
\ee
In the absence of disclinations the momentum current, $\sigma^i_A$, has the meaning of dislocation current. The momentum density, $P_A$, is proportional to the Burgers vector of the background dislocation, but is only a \emph{fraction} by the magnitude. The smallest possible Burgers vector is determined by the primitive lattice vector of the underlying lattice. The theory \eqref{eq:FQEsource} can be visualized as a fluctuating quantum lattice, described geometrically in terms of $\hat e^A_\mu$, that is formed in a quantum system with the underlying classical ``ion'' lattice, described geometrically by the classical sources $e^A_\mu$. We can probe the quantum elastic system by distorting the classical  lattice (see Fig \ref{mom}).

%%%%%%%%%%%%
\paragraph{Glide constraint.---}
%%%%%%%%%%%%

Fracton phases, as well as the present theory of topological elasticity, exhibit excitations with restricted mobility. In the classical elasticity theory this phenomenon is referred to as the \emph{glide constraint}  \cite{marchetti1999interstitials, cvetkovic2006topological} --- a dislocation can only move (or glide) in the direction of their Burgers vector. The motion perpendicular to the Burgers vector (or climb) requires adding an interstitial or a vacancy. At low temperatures, the density of interstitials is very low and climb is prohibited. The present theory also satisfies the glide constraint. To see it we take the inspiration from \cite{cvetkovic2006topological,beekman2017dual} and define the transverse dislocation current 
\be\la{eq:sigmaperp}
\sigma_\perp = \epsilon^A{}_B \hat e^B_i \sigma^i_A  = -\frac{1}{\pi} \left[ \frac{1}{2} (\p_0 \hat e) - \epsilon_{AB} \epsilon^{ij} \hat e^B_i \p_j \hat e^{A}_0 \right]\,,
\ee 
which has the meaning of the dislocation current perpendicular to the Burgers vector. In the notations of Eq.\,\eqref{eq:currents} $\sigma_\perp \propto b_A  \epsilon^A{}_B \dot{a}^B = \mathbf{b} \times \dot{\mathbf{a}}$.
The total mass (or, total number of lattice sites) is defined as  the total volume $M = \int d^2 x \hat e$ of the quantum elastic medium \cite{katanaev2005geometric}. In the absence of mobile interstitials or vacancies the local volume should be conserved. Such conservation is precisely the  expression in square brackets in \eqref{eq:sigmaperp} \footnote{Vanishing of the r.h.s. of \eqref{eq:sigmaperp} is an identity if we consider the non-singular , holonomic, vielbein $\hat e^A_\mu = \p_\mu u^A$, where $u^A$ is the distortion vector. Since vacancies and interstitial are topologically trivial this check is sufficient.}, but then $\sigma_\perp =0$. Thus local, non-linear conservation of mass (or volume) implies that dislocations can only move along their Burgers vectors. This constraint reduces to a similar relationship discussed in the elasticity literature \cite{marchetti1999interstitials,cvetkovic2006topological} upon linearization.

%%%%%%%%%%%%%%%%%%%%%%
\paragraph{Disclinations, Curvature and Fractons.---} 
%%%%%%%%%%%%%%%%%%%%%%

From the point of view of elasticity theory the disclinations are very high energy defects since they require a removal (or addition) of a macroscopic amount of material. These defects are described by the singular configuration of the (dynamic) curvature $\hat R$, which can be described by the torsion alone by the virtue of Cartan structure equations
\be\la{eq:cart}
\hat T^A = d \hat e^A + \epsilon^A{}_B \hat \omega \wedge \hat e^B\,.
\ee
These equations should be viewed as the definition of the spin connection $\hat \omega_\mu$, given the vielbeins and torsion \cite{bradlyn2014low,gromov-thermal}.
Then the conservation of momentum current, $\p_\mu \sigma_A^\mu \equiv 0$, leads to a relation between the curvature and torsion, known as the Bianchi identity. This relationship becomes physically transparent when we define the current of dislocations and disclinations according to
\be
J^\mu_{A} = \epsilon^{\mu\nu\rho} \hat T_{A,\nu\rho}\,, \qquad \Theta^\mu = \epsilon^{\mu\nu\rho}\hat R_{\nu\rho}\,,
\ee
 where $\hat R_{\mu\nu} = 2(\p_\mu \hat \omega_\nu - \p_\nu \hat \omega_\mu)$ is the curvature $2$-form. Then the Bianchi identity takes form
\be\la{eq:discldisl}
\nabla_\mu J^\mu_A = \epsilon_{AB} \hat e^B_\rho \Theta^\rho\,,
\ee
where $\nabla_\mu J^\mu_A = \p_\mu J^\mu_A - \hat \omega_\mu \epsilon^B{}_AJ^\mu_B$ is the covariant divergence. Eq.\,\eqref{eq:discldisl} is a non-linear, covariant generalization of the usual relation between dislocations and disclinations: it tells us that disclination current, $\Theta^i$, \emph{must} be accompanied by creation of dislocations with Burgers vector perpendicular to $\Theta^i$. 

It turns out that the density of dislocations and disclinations are related. Consider a purely spatial, $2$D, version of \eqref{eq:cart}. One possible choice of the spin connection, known as the Levi-Civita connection, is the one that corresponds to the vanishing torsion. We denote it as $\hat \omega_i^{\rm LC}$. The general connection can be written as $\hat \omega_i = \hat \omega_i^{\rm LC} + \hat C_i$, where $\hat C_i = \epsilon_{ij} \epsilon^A{}_B e^j_A \hat T^B$ is the contorsion \cite{hughes2012torsion}. We now specify to the teleparallel (curvature-free) case $\hat \omega_i \equiv 0$. Then
\be\la{eq:GB}
\hat \omega_i^{\rm LC}  =  \epsilon_{ij} \epsilon_A{}^B e^j_B \hat T^A \quad \Rightarrow \quad 2\hat R^{\rm LC} = \p_i \Big(\epsilon^A{}_B \hat e^i_A \hat T^B\Big)\,,
\ee
where $\hat R^{\rm LC}$ is the curvature of $\hat \omega^{\rm LC}$.
Mathematical consequence of \eqref{eq:GB} is the equivalence between curvature-free and torsion-free descriptions of the $2$D geometry \footnote{In general relativity the torsion-free approach is usually taken, whereas in the theory of elasticity the curvature-free perspective is more convenient since dislocations are the low energy objects.}. In particular, Gauss-Bonnet theorem can be phrased in terms of either curvature or torsion. Physical consequence of this relation is the identification of a disclination dipole with a dislocation, whose Burgers vector is perpendicular to the dipole. The glide constraint then implies that the disclination dipole can only move perpendicular to its dipole moment. 
These relations are another piece of the Fracton phenomenology: disclinations are immobile excitations, that can only move by exchanging the dislocations. The latter can be regarded as disclination dipoles. 

An important comment is in order: regardless of the relation to elasticity, the non-linear identities of the Riemann-Cartan geometry (or the gauge theory of $\mathbb R^2 \rtimes SO(2)$) contain some of the phenomenology of the Fracton order. This suggests that the Fracton order itself may be of \emph{geometric} origin.

%%%%%%%%%%%%%%%%%%%%%%
\paragraph{Linear response.---} 
%%%%%%%%%%%%%%%%%%%%%%

Integrating out the quantum fields $\hat e^A_\mu$, and neglecting the global issues, we find the generating functional and  momentum current response to the time-dependent variation of background geometry
\be
W\left[e^A_\mu\right] = \frac{\delta_{AB}}{4\pi k} \int e^A \wedge d e^B \quad \Rightarrow \quad \left\langle \sigma_A^i \right\rangle = \frac{1}{2\pi k}  \epsilon^{ij}  \p_0e_{A,j}\,.
\ee
This response is known as ``torsional Hall viscosity'', studied for Chern insulators \cite{hughes2011torsional,hughes2012torsion, geracie2014hall,hoyos2014hall}, and is identical to the ``generalized'' Hall response of \cite{prem2017emergent} upon linearization. There is an important difference from the traditional electromagnetic Hall response. The generating functional $W\left[e^A_\mu\right]$ is \emph{locally} invariant (\emph{i.e.}\,does not transform by a total derivative) under all symmetries of the problem. Thus, if we were to introduce a boundary, it will not require a ``compensating'' anomalous gapless degree of freedom \footnote{This is also true for the usual Hall viscosity \cite{gromov2016boundary}}. Consequently, we do not expect a robust edge mode. This conclusion does not immediately contradict the (opposite) results discussed in \cite{prem2017emergent}. It is possible that some microscopic models, with a particular choice of boundary conditions will support gapless edge modes.

%%%%%%%%%%%%
\paragraph{Conclusions.---}
%%%%%%%%%%%%
We have found that the higher rank gauge theories are no longer gauge invariant if arbitrarily small and smooth curvature is introduced. We have proposed an alternative effective theory that does not suffer such problem, because it does not possess a gauge ``symmetry''. This theory describes chiral topological elasticity and exhibits the phenomenology of the Fracton models. The realm of topological (or global) properties of the (chiral) quantum elasticity appears to be largely, if not completely, unexplored. It would be very interesting to develop the canonical quantization of the topological elasticity on a torus, understand the relationship to the gauge theory of translations, derive the sub-extensive groundstate degeneracy, directly show the instability of the edge modes and identify the fractal operators present in some Fracton models \cite{haah2011local, yoshida2013exotic}.

%%%%%%%%%%
\acknowledgements
%%%%%%%%%%

I thank Vincenzo Vitelli and especially Barry Bradlyn for the very insightful discussions.
A.G. was supported by the Leo Kadanoff fellowship and by the University of Chicago Materials Research Science and Engineering Center, which is funded by the National Science Foundation under award number DMR-1420709.

%%%%%%%%%%
\paragraph{Note added.---}When the present manuscript was in preparation I have learned about Ref.\,\cite{PretkoPRL2018}, where the relation of Fracton models to the elasticity was discussed and identification of lattice defects with Fractons was made. We have also became aware of \cite{slagleXcube2017}, where the geometric nature of the Fracton order was emphasized, through a different line of reasoning.
%%%%%%%%%%

\bibliography{Bibliography}

\newpage

\end{document}